\documentclass[seceq,supplement]{ptptex}

\title{Naturalness and Testability of TeV Seesaw Mechanisms}

\author{Zhi-zhong \textsc{Xing}\footnote{E-mail: xingzz@ihep.ac.cn}}

\inst{Institute of High Energy Physics and Theoretical Physics
Center for Science Facilities, Chinese Academy of Sciences, Beijing
100049, China}

\abst{After outlining some popular ways to go beyond the standard
model so as to generate non-zero but tiny neutrino masses, I focus
on several typical seesaw mechanisms and discuss how to get a
balance between their theoretical naturalness and their
experimental testability. Besides possible collider signatures at
the Large Hadron Collider, new and non-unitary CP-violating
effects are also expected to show up in neutrino oscillations for
type-I, type-(I+II), type-III and double seesaws at the TeV
scale.}

\begin{document}

\maketitle

\section{Possible ways to go beyond the SM}

The Large Hadron Collider (LHC) is presumably bringing us to a new
energy frontier --- the TeV scale, at which some fundamental new
physics beyond the standard model (SM) of electroweak interactions
is expected to show up and reveal the origin of masses of elementary
particles. Can the LHC help us to understand the origin of neutrino
masses? In other words, could TeV neutrino physics become an
exciting direction in the LHC era? This question is just the
motivation of my talk.

Neutrinos are assumed or required to be massless in the SM, just
because the structure of the SM itself is too simple to
accommodate massive neutrinos.
\begin{itemize}
\item     Two fundamentals of the SM are the $SU(2)^{}_{\rm L} \times
U(1)^{}_{\rm Y}$ gauge symmetry and the Lorentz invariance. Both of
them are mandatory to guarantee that the SM is a consistent quantum
field theory.

\item     The particle content of the SM is rather economical. There
are no right-handed neutrinos in the SM, so a Dirac neutrino mass
term is not allowed. There is only one Higgs doublet, so a
gauge-invariant Majorana mass term is forbidden.

\item     The SM is a renormalizable quantum field theory. Hence
an effective dimension-5 operator, which may give each neutrino a
Majorana mass, is absent.
\end{itemize}
In other words, the SM accidently possesses the $(B-L)$ symmetry
which assures three known neutrinos to be exactly massless.

But today's experiments have convincingly indicated the existence
of neutrino oscillations \cite{PDG}. This quantum phenomenon can
appear if and only if neutrinos are massive and lepton flavors are
mixed, and thus it is a kind of new physics beyond the SM. To
generate non-zero but tiny neutrino masses, one or more of the
above-mentioned constraints on the SM must be abandoned or
relaxed. It is certainly intolerable to abandon the gauge symmetry
and Lorentz invariance \cite{Feruglio}; otherwise, one would be
led astray. Given the framework of the SM as a consistent field
theory, its particle content can be modified and (or) its
renormalizability can be abandoned to accommodate massive
neutrinos. There are several ways to this goal.

\subsection{To relax the requirement of renormalizability}

In 1979, Weinberg \cite{W} extended the SM by introducing some
higher-dimension operators in terms of the fields of the SM itself:
\begin{equation}
{\cal L}^{}_{\rm eff} \; = \; {\cal L}^{}_{\rm SM} + \frac{{\cal
L}^{}_{\rm d=5}}{\Lambda} + \frac{{\cal L}^{}_{\rm d=6}}{\Lambda^2}
+ \cdots \; ,
\end{equation}
where $\Lambda$ denotes the cut-off scale of this effective
theory. Within such a framework, the lowest-dimension operator
that violates the lepton number ($L$) is the unique dimension-5
operator $HHLL/\Lambda$. After spontaneous gauge symmetry
breaking, this Weinberg operator yields $m^{}_i \sim \langle
H\rangle^2/\Lambda$ for neutrino masses, which can be sufficiently
small ($\lesssim 1$ eV) if $\Lambda$ is not far away from the
scale of grand unified theories ($\Lambda \gtrsim 10^{13}$ GeV for
$\langle H\rangle \sim 10^2$ GeV). In this sense, people argue
that neutrino masses can serve as a low-energy window onto new
physics at superhigh energy scales.

\subsection{To add three right-handed neutrinos and demand $(B-L)$
symmetry}

Given three right-handed neutrinos, the gauge-invariant and
lepton-number-conserving mass terms of charged leptons and
neutrinos are
\begin{equation}
-{\cal L}^{}_{\rm lepton} = \overline{l^{}_{\rm L}} Y^{}_l H
E^{}_{\rm R} + \overline{l^{}_{\rm L}} Y^{}_\nu \tilde{H}
N^{}_{\rm R} + {\rm h.c.} \; ,
\end{equation}
where $\tilde{H} \equiv i\sigma^{~}_2 H^*$ is defined and $l_{\rm
L}$ denotes the left-handed lepton doublet. After spontaneous
gauge symmetry breaking, we arrive at the charged-lepton mass
matrix $M^{}_l = Y^{}_l v/\sqrt{2}$ and the Dirac neutrino mass
matrix $M^{}_\nu = Y^{}_\nu v/\sqrt{2}$ with $v \simeq 246 ~ {\rm
GeV}$. In this case, the smallness of three neutrino masses
$m^{}_i$ (for $i=1,2,3$) is attributed to the smallness of three
eigenvalues of $Y^{}_\nu$ (denoted as $y^{}_i$ for $i=1,2,3$).
Then we encounter a transparent hierarchy problem: $y^{}_i/y^{}_e
= m^{}_i/m^{}_e \lesssim 0.5 ~{\rm eV}/0.5 ~{\rm MeV} \sim
10^{-6}$. Why is $y^{}_i$ so small? There is no explanation at all
in this Dirac-mass picture.

A speculative way out is to invoke extra dimensions; namely, the
smallness of Dirac neutrino masses is ascribed to the assumption
that three right-handed neutrinos have access to one or more extra
spatial dimensions \cite{ED}. The idea is simply to confine the SM
particles onto a brane and to allow $N^{}_{\rm R}$ to travel in
the bulk. For example, the wave-function of $N^{}_{\rm R}$ spreads
out over the extra dimension $y$, giving rise to a suppressed
Yukawa interaction at $y = 0$ (i.e., the location of the brane):
\begin{equation}
\left[ \overline{l^{}_{\rm L}} Y^{}_\nu \tilde{H} N^{}_{\rm R}
\right]^{}_{y=0} ~ \sim ~ \frac{1}{\sqrt{L}} \left[
\overline{l^{}_{\rm L}} Y^{}_\nu \tilde{H} N^{}_{\rm R}
\right]^{}_{y=L} \; .
\end{equation}
The magnitude of $1/\sqrt{L}$ is measured by
$\Lambda/\Lambda^{}_{\rm Planck}$, and thus it can naturally be
small for an effective theory far below the Planck scale.

\subsection{To add new degrees of freedom and allow $(B-L)$
violation}

This approach works at the tree level and reflects the essential
spirit of seesaw mechanisms --- tiny masses of three known
neutrinos are attributed to the existence of heavy degrees of
freedom and lepton number violation. There are three simple and
typical seesaw mechanisms on the market.
\begin{itemize}
\item     Type-I seesaw --- three heavy right-handed neutrinos are
added into the SM and the lepton number is violated by their
Majorana mass term \cite{T1SS}:
\begin{equation}
-{\cal L}^{}_{\rm lepton} \; = \; \overline{l^{}_{\rm L}} Y^{}_l H
E^{}_{\rm R} + \overline{l^{}_{\rm L}} Y^{}_\nu \tilde{H}
N^{}_{\rm R} + \frac{1}{2} \overline{N^{\rm c}_{\rm R}} M^{}_{\rm
R} N^{}_{\rm R} + {\rm h.c.} \; ,
\end{equation}
where $M^{}_{\rm R}$ is the symmetric Majorana mass matrix.

\item     Type-II seesaw --- one heavy Higgs triplet is added into the
SM and the lepton number is violated by its interactions with both
the lepton doublet and the Higgs doublet \cite{T2SS}:
\begin{equation}
-{\cal L}^{}_{\rm lepton} \; = \; \overline{l^{}_{\rm L}} Y^{}_l H
E^{}_{\rm R} + \frac{1}{2} \overline{l^{}_{\rm L}} Y^{}_\Delta
\Delta i\sigma^{}_2 l^{\rm c}_{\rm L} - \lambda^{}_\Delta
M^{}_\Delta H^T i\sigma^{}_2 \Delta H + {\rm h.c.} \; ,
\end{equation}
where \begin{equation} \Delta \; \equiv \; \left(\begin{matrix}
\Delta^- & -\sqrt{2} ~ \Delta^0 \cr \sqrt{2} ~ \Delta^{--} &
-\Delta^- \cr \end{matrix} \right)
\end{equation}
denotes the $SU(2)^{}_{\rm L}$ Higgs triplet.

\item     Type-III seesaw --- three heavy triplet fermions are
added into the SM and the lepton number is violated by their
Majorana mass term \cite{T3SS}:
\begin{equation}
-{\cal L}^{}_{\rm lepton} \; = \; \overline{l^{}_{\rm L}} Y^{}_l H
E^{}_{\rm R} + \overline{l^{}_{\rm L}} \sqrt{2} Y^{}_\Sigma
\Sigma^{\rm c} \tilde{H} + \frac{1}{2} {\rm Tr} \left(
\overline{\Sigma} M^{}_\Sigma \Sigma^{\rm c} \right) + {\rm h.c.} \;
,
\end{equation}
where
\begin{equation}
\Sigma \; =\; \left( \begin{matrix} \Sigma^0/\sqrt{2} & \Sigma^+
\cr \Sigma^- & -\Sigma^0/\sqrt{2} \cr\end{matrix} \right)
\end{equation}
denotes the $SU(2)^{}_{\rm L}$ fermion triplet.
\end{itemize}
Of course, there are a number of variations or combinations of
these three typical seesaw mechanisms in the literature.

For each of the above seesaw pictures, one may arrive at the
unique dimension-5 Weinberg operator of neutrino masses after
integrating out the corresponding heavy degrees of freedom
\cite{Zhou}:
\begin{equation}
\frac{{\cal L}^{}_{\rm d=5}}{\Lambda} \; =\; \left\{
\begin{array}{lcl}
\displaystyle \frac{1}{2} \left(Y^{}_\nu M^{-1}_{\rm R}
Y^T_\nu\right)^{}_{\alpha\beta} \overline{l^{}_{\alpha \rm L}}
\tilde{H} \tilde{H}^T l^{\rm c}_{\beta \rm L} + {\rm h.c.} && ({\rm
Type ~I}) \; , \vspace{0.15cm} \\
\displaystyle -\frac{\lambda^{}_\Delta}{M^{}_\Delta}
\left(Y^{}_\Delta\right)^{}_{\alpha\beta} \overline{l^{}_{\alpha \rm
L}} \tilde{H} \tilde{H}^T l^{\rm c}_{\beta \rm L} + {\rm h.c.} &&
({\rm Type ~II}) \; , \vspace{0.15cm} \\
\displaystyle \frac{1}{2}
\left(Y^{}_\Sigma M^{-1}_{\rm \Sigma}
Y^T_\Sigma\right)^{}_{\alpha\beta} \overline{l^{}_{\alpha \rm L}}
\tilde{H} \tilde{H}^T l^{\rm c}_{\beta \rm L} + {\rm h.c.} && ({\rm
Type ~III}) \; .
\end{array}
\right .
\end{equation}
After spontaneous gauge symmetry breaking, $\tilde{H}$ achieves
its vacuum expectation value $\langle \tilde{H}\rangle =
v/\sqrt{2}$ with $v \simeq 246$ GeV. Then we are left with the
effective Majorana neutrino mass term for three known neutrinos,
\begin{equation}
-{\cal L}^{}_{\rm mass} \; =\; \frac{1}{2} \overline{\nu^{}_{\rm L}}
M^{}_\nu \nu^{\rm c}_{\rm L} + {\rm h.c.} \; ,
\end{equation}
where the symmetric Majorana mass matrix $M^{}_\nu$ is given by
\begin{equation}
M^{}_\nu \; =\; \left\{
\begin{array}{lcl}
\displaystyle -\frac{1}{2} Y^{}_\nu \frac{v^2}{M^{}_{\rm R}} Y^T_\nu
&& ({\rm Type ~I}) \; , \vspace{0.15cm} \\
\displaystyle
\lambda^{}_\Delta Y^{}_\Delta \frac{v^2}{M^{}_\Delta } && ({\rm Type
~II}) \; , \vspace{0.15cm} \\
\displaystyle -\frac{1}{2} Y^{}_\Sigma
\frac{v^2}{M^{}_\Sigma} Y^T_\Sigma && ({\rm Type ~III}) \; .
\end{array}
\right .
\end{equation}
It becomes obvious that the smallness of $M^{}_\nu$ can be
attributed to the largeness of $M^{}_{\rm R}$, $M^{}_\Delta$ or
$M^{}_\Sigma$ in the seesaw mechanism.

\subsection{Radiative generation of tiny neutrino masses}

In a seminal paper published in 1972, Weinberg \cite{W1972}
pointed out that ``in theories with spontaneously broken gauge
symmetries, various masses or mass differences may vanish in
zeroth order as a consequence of the representation content of the
fields appearing in the Lagrangian. These masses or mass
differences can then be calculated as finite higher-order
effects." Such a mechanism may allow us to slightly go beyond the
SM and radiatively generate tiny neutrino masses. A typical
example is the well-known Zee model \cite{Zee},
\begin{equation}
-{\cal L}^{}_{\rm lepton} \; = \; \overline{l^{}_{\rm L}} Y^{}_l H
E^{}_{\rm R} + \overline{l^{}_{\rm L}} Y^{}_S S^- i\sigma^{}_2
l^{\rm c}_{\rm L} + \tilde{\Phi}^T F S^+ i\sigma^{}_2 \tilde{H} +
{\rm h.c.} \; ,
\end{equation}
where $S^{\pm}$ are charged $SU(2)^{}_{\rm L}$ singlet scalars,
$\Phi$ denotes a new $SU(2)^{}_{\rm L}$ doublet scalar which has the
same quantum number as the SM Higgs doublet $H$, $Y^{}_S$ is an
anti-symmetric matrix, and $F$ represents a mass. Without loss of
generality, we choose the basis of $M^{}_l = Y^{}_l \langle H\rangle
= {\rm Diag}\{m^{}_e, m^{}_\mu, m^{}_\tau \}$. In this model,
neutrinos are massless at the tree level, but their masses can
radiatively be generated via the one-loop corrections. Given $M^{}_S
\gg M^{}_H \sim M^{}_\Phi \sim F$ and $\langle \Phi\rangle \sim
\langle H\rangle$, the elements of the effective mass matrix of
three light Majorana neutrinos turns out to be
\begin{equation}
\left(M^{}_\nu\right)^{}_{\alpha\beta} \; \sim \;
\frac{M^{}_H}{16\pi^2} \cdot \frac{m^2_\alpha - m^2_\beta}{M^2_S}
\left(Y^{}_S\right)^{}_{\alpha\beta} \; ,
\end{equation}
where $\alpha$ and $\beta$ run over $e$, $\mu$ and $\tau$. The
smallness of $M^{}_\nu$ is therefore ascribed to the smallness of
$Y^{}_S$ and $(m^2_\alpha - m^2_\beta)/M^2_S$. Although the
original version of the Zee model is disfavored by current
experimental data on neutrino oscillations, its extensions or
variations at the one-loop or two-loop level can survive
\cite{Babu}.

\section{On the scale of seesaw mechanisms}

As we have seen, the key point of a seesaw mechanism is to ascribe
the smallness of neutrino masses to the existence of some new
degrees of freedom heavier than the Fermi scale $v \simeq 246$
GeV, such as heavy Majorana neutrinos or heavy Higgs bosons. The
energy scale where a seesaw mechanism works is crucial, because it
is relevant to whether this mechanism is theoretically natural and
experimentally testable. Between Fermi and Planck scales, there
might exist two other fundamental scales: one is the scale of a
grand unified theory (GUT) at which strong, weak and
electromagnetic forces can be unified, and the other is the TeV
scale at which the unnatural gauge hierarchy problem of the SM can
be solved or at least softened by a kind of new physics.

\subsection{How about a very low seesaw scale?}

In reality, however, there is no direct evidence for a high or
extremely high seesaw scale. Hence eV-, keV-, MeV- and GeV-scale
seesaws are all possible, at least in principle, and they are
technically natural in the sense that their
lepton-number-violating mass terms are naturally small according
to 't Hooft's naturalness criterion \cite{Hooft}
--- ``At any energy scale $\mu$, a set of parameters $\alpha^{}_i
(\mu)$ describing a system can be small, if and only if, in the
limit $\alpha^{}_i (\mu) \rightarrow 0$ for each of these
parameters, the system exhibits an enhanced symmetry." But there
are several potential problems associated with low-scale seesaws
\cite{Gouvea}: (a) a low-scale seesaw does not give any obvious
connection to a theoretically well-justified fundamental physical
scale (such as the Fermi scale, the TeV scale, the GUT scale or
the Planck scale); (b) the neutrino Yukawa couplings in a
low-scale seesaw model turn out to be tiny, giving no actual
explanation of why the masses of three known neutrinos are so
small; and (c) in general, a very low seesaw scale does not allow
the ``canonical" thermal leptogenesis mechanism \cite{FY} to work,
although there might be a way out.

\subsection{The hierarchy problem of conventional seesaws}

Many theorists argue that the conventional seesaw scenarios are
natural because their scales (i.e., the masses of heavy degrees of
freedom) are close to the GUT scale. This argument is reasonable
on the one hand, but it reflects the drawbacks of the conventional
seesaw models on the other hand. In other words, the conventional
seesaw models have no direct experimental testability and involve
a potential hierarchy problem. The latter is usually spoke of when
two largely different energy scales exist in a model, but there is
no symmetry to stabilize the low-scale physics suffering from
large corrections coming from the high-scale physics.

Such a seesaw-induced fine-tuning problem means that the SM Higgs
mass is very sensitive to quantum corrections from the heavy degrees
of freedom in a seesaw mechanism. For example
\cite{Vissani98,Abada},
\begin{equation}
\delta M^2_H \; = \; \left\{
\begin{array}{lcl}
\displaystyle -\frac{y^2_i}{8\pi^2} \left(\Lambda^2 + M^2_i
\ln\frac{M^2_i}{\Lambda^2} \right) && ({\rm Type ~I}) \vspace{0.2cm} \\
\displaystyle \frac{3}{16\pi^2} \left[ \lambda^{}_3 \left(\Lambda^2
+ M^2_\Delta \ln\frac{M^2_\Delta}{\Lambda^2}\right) + 4
\lambda^2_\Delta M^2_\Delta \ln\frac{M^2_\Delta}{\Lambda^2} \right]
\hspace{-0.2cm} && ({\rm Type ~II}) \vspace{0.2cm} \\ \displaystyle
-\frac{3y^2_i}{8\pi^2} \left(\Lambda^2 + M^2_i
\ln\frac{M^2_i}{\Lambda^2} \right) && ({\rm Type ~III})
\end{array}
\right .
\end{equation}
in three typical seesaw scenarios, where $\Lambda$ is the regulator
cut-off, $y^{}_i$ and $M^{}_i$ (for $i=1,2,3$) stand respectively
for the eigenvalues of $Y^{}_\nu$ (or $Y^{}_\Sigma$) and $M^{}_{\rm
R}$ (or $M^{}_\Sigma$), and the contributions proportional to $v^2$
and $M^2_H$ have been omitted. Eq. (2.1) show a quadratic
sensitivity to the new scale which is characteristic of the seesaw
model, implying that a high degree of fine-tuning would be necessary
to accommodate the experimental data on $M^{}_H$ if the seesaw scale
is much larger than $v$ (or the Yukawa couplings are not extremely
fine-tuned in type-I and type-III seesaws) \cite{Abada}.

Taking the type-I seesaw scenario for illustration, we assume
$\Lambda \sim M^{}_i$ and require $|\delta M^2_H| \lesssim 0.1
~{\rm TeV}^2$. Then Eq. (2.1) leads us to the following rough
estimate:
\begin{equation}
M^{}_i \; \sim \; \left[\frac{(2\pi v)^2 |\delta
M^2_H|}{m^{}_i}\right]^{1/3} \lesssim 10^7 {\rm GeV} \left[\frac{0.2
~ {\rm eV}}{m^{}_i}\right]^{1/3} \left[\frac{|\delta M^2_H|}{0.1 ~
{\rm TeV}^2}\right]^{1/3} \; .
\end{equation}
This naive result indicates that a hierarchy problem will arise if
the masses of heavy Majorana neutrinos are larger than about $10^7$
GeV in the type-I seesaw scheme. Because of $m^{}_i \sim y^2_i
v^2/(2M^{}_i)$, the bound $M^{}_i \lesssim 10^7$ GeV implies $y^{}_i
\sim \sqrt{2m^{}_i M^{}_i}/v \lesssim 2.6 \times 10^{-4}$ for
$m^{}_i \sim 0.2$ eV. Such a small magnitude of $y^{}_i$ seems to be
a bit unnatural in the sense that the conventional seesaw idea
attributes the smallness of $m^{}_i$ to the largeness of $M^{}_i$
other than the smallness of $y^{}_i$.

There are two possible ways out of this impasse: one is to appeal
for the supersymmetry, and the other is to lower the seesaw scale.
In the remaining part of this talk, I shall follow the second way to
discuss the TeV seesaw mechanisms which do not suffer from the
above-mentioned hierarchy problem.

\subsection{Why are the TeV seesaws interesting?}

There are several reasons for people to expect some new physics at
the TeV scale. This kind of new physics should be able to
stabilize the Higgs-boson mass and hence the electroweak scale; in
other words, it should be able to solve or soften the unnatural
gauge hierarchy problem. It has also been argued that the
weakly-interacting particle candidates for dark matter should
weigh about one TeV or less \cite{Dimopoulos}. If the TeV scale is
really a fundamental scale, may we argue that the TeV seesaws are
natural? Indeed, we are reasonably motivated to speculate that
possible new physics existing at the TeV scale and responsible for
the electroweak symmetry breaking might also be responsible for
the origin of neutrino masses \cite{ICHEP08}. It is interesting
and meaningful in this sense to investigate and balance the
``naturalness" and ``testability" of TeV seesaws at the energy
frontier set by the LHC.

As a big bonus of the conventional (type-I) seesaw mechanism, the
thermal leptogenesis mechanism \cite{FY} provides us with an
elegant dynamic picture to interpret the cosmological
matter-antimatter asymmetry characterized by the observed ratio of
baryon number density to photon number density, $\eta^{}_B \equiv
n^{}_B/n^{}_\gamma = (6.1 \pm 0.2) \times 10^{10}$. When heavy
Majorana neutrino masses are down to the TeV scale, the Yukawa
couplings should be reduced by more than six orders of magnitude
so as to generate tiny masses for three known neutrinos via the
type-I seesaw and satisfy the out-of-equilibrium condition, but
the CP-violating asymmetries of heavy Majorana neutrino decays can
still be enhanced by the resonant effects in order to account for
$\eta^{}_B$. This ``resonant leptogenesis" scenario might work in
a specific TeV seesaw model \cite{Pilaftsis}.

Is there a TeV Noah's Ark which can naturally and simultaneously
accommodate the seesaw idea, the leptogenesis picture and the
collider signatures? We are most likely not so lucky and should not
be too ambitious at present. In the subsequent sections, we shall
concentrate on the TeV seesaws themselves and their possible
collider signatures and low-energy consequences.

\section{Naturalness and testability of TeV seesaws}

The neutrino mass terms in three typical seesaw mechanisms have been
given in section $\S$1.3. Without loss of generality, we choose the
basis in which the mass eigenstates of three charged leptons are
identified with their flavor eigenstates.

\subsection{Type-I seesaw}

Given $M^{}_{\rm D} = Y^{}_\nu v/\sqrt{2}~$, the approximate
type-I seesaw formula in Eq. (1.11) can be rewritten as $M^{}_\nu
= - M^{}_{\rm D} M^{-1}_{\rm R} M^T_{\rm D}$. Note that the
$3\times 3$ light neutrino mixing matrix $V$ is not exactly
unitary in this seesaw scheme, and its deviation from unitarity is
of ${\cal O}(M^2_{\rm D}/M^2_{\rm R})$. Let us consider two
interesting possibilities.
\begin{itemize}
\item     $M^{}_{\rm D} \sim {\cal O}(10^2)$ GeV and $M^{}_{\rm R} \sim
{\cal O}(10^{15})$ GeV to get $M^{}_\nu \sim {\cal O}(10^{-2})$ eV.
In this conventional and {\it natural} case, $M^{}_{\rm D}/M^{}_{\rm
R} \sim {\cal O}(10^{-13})$ holds. Hence the non-unitarity of $V$ is
only at the ${\cal O}(10^{-26})$ level, too small to be observed.

\item     $M^{}_{\rm D} \sim {\cal O}(10^2)$ GeV and $M^{}_{\rm R} \sim
{\cal O}(10^{3})$ GeV to get $M^{}_\nu \sim {\cal O}(10^{-2})$ eV.
In this {\it unnatural} case, a significant ``structural
cancellation" has to be imposed on the textures of $M^{}_{\rm D}$
and $M^{}_{\rm R}$. Because of $M^{}_{\rm D}/M^{}_{\rm R} \sim {\cal
O}(0.1)$, the non-unitarity of $V$ can reach the percent level and
may lead to some observable effects.
\end{itemize}
Now we discuss how to realize the above ``structural cancellation"
for the type-I seesaw mechanism at the TeV scale. For the sake of
simplicity, we take the basis of $M^{}_{\rm R} = {\rm
Diag}\{M^{}_1, M^{}_2, M^{}_3\}$ for three heavy Majorana
neutrinos ($N^{}_1, N^{}_2, N^{}_3$). It is well known that
$M^{}_\nu$ vanishes if
\begin{equation}
M^{}_{\rm D} \; = \; m \left( \begin{matrix} y^{}_1 & y^{}_2 &
y^{}_3 \cr \alpha y^{}_1 & \alpha y^{}_2 & \alpha y^{}_3 \cr \beta
y^{}_1 & \beta y^{}_2 & \beta y^{}_3 \cr \end{matrix} \right) ~~~~~~
{\rm and} ~~~~~~~~ \sum^3_{i=1} \frac{y^2_i}{M^{}_i} \; = \; 0
\end{equation}
simultaneously hold \cite{Xing09}. Tiny neutrino masses can be
generated from tiny corrections to the texture of $M^{}_{\rm D}$
in Eq. (3.1). For example, $M^\prime_{\rm D} = M^{}_{\rm D} -
\epsilon X^{}_{\rm D}$ with $M^{}_{\rm D}$ given above and
$\epsilon$ being a small dimensionless parameter (i.e.,
$|\epsilon| \ll 1$) will yield
\begin{equation}
M^\prime_\nu \; = \; -M^\prime_{\rm D} M^{-1}_{\rm R} M^{\prime
T}_{\rm D} \; \simeq \; \epsilon \left( M^{}_{\rm D} M^{-1}_{\rm
R} X^T_{\rm D} + X^{}_{\rm D} M^{-1}_{\rm R} M^T_{\rm D} \right)
\; ,
\end{equation}
from which $M^\prime_\nu \sim {\cal O}(10^{-2})$ eV can be
obtained by adjusting the size of $\epsilon$.

A lot of attention has recently been paid to a viable type-I seesaw
model and its collider signatures at the TeV scale \cite{ICHEP08}.
At least the following lessons can be learnt:
\begin{itemize}
\item     Two necessary conditions must be satisfied in order to
test a type-I seesaw model at the LHC: (a) $M^{}_i$ are of ${\cal
O}(1)$ TeV or smaller; and (b) the strength of light-heavy
neutrino mixing (i.e., $M^{}_{\rm D}/M^{}_{\rm R}$) is large
enough. Otherwise, it would be impossible to produce and detect
$N^{}_i$ at the LHC.

\item     The collider signatures of $N^{}_i$ are essentially
decoupled from the mass and mixing parameters of three light
neutrinos $\nu^{}_i$. For instance, the small parameter $\epsilon$
in Eq. (3.2) has nothing to do with the ratio $M^{}_{\rm
D}/M^{}_{\rm R}$.

\item     The non-unitarity of $V$ might lead to some
observable effects in neutrino oscillations and other
lepton-flavor-violating or lepton-number-violating processes, if
$M^{}_{\rm D}/M^{}_{\rm R} \lesssim {\cal O}(0.1)$ holds. More
discussions will be given in section $\S$4.

\item     The clean LHC signatures of heavy Majorana neutrinos are
the $\Delta L =2$ like-sign dilepton events \cite{Senjanovic}, such
as $pp \to W^{*\pm} W^{*\pm} \to \mu^\pm \mu^\pm jj$ (a collider
analogue to the neutrinoless double-beta decay) and $pp \to W^{*\pm}
\to \mu^\pm N_i \to \mu^\pm \mu^\pm jj$ (a dominant channel due to
the resonant production of $N^{}_i$).
\end{itemize}
Some instructive and comprehensive analyses of possible LHC events
for a single heavy Majorana neutrino have recently been done
\cite{Han}, but they only serve for illustration because such a
simplified type-I seesaw scenario is actually unrealistic.

\subsection{Type-II seesaw}

The type-II seesaw formula $M^{}_\nu = Y^{}_\Delta v^{}_\Delta =
\lambda^{}_\Delta Y^{}_\Delta v^2/M^{}_\Delta$ has already been
given in Eq. (1.11). Note that the last term of Eq. (1.5) violates
both $L$ and $B-L$, and thus the smallness of $\lambda^{}_\Delta$
is naturally allowed according to 't Hooft's naturalness criterion
(i.e., setting $\lambda^{}_\Delta =0$ will increase the symmetry
of ${\cal L}^{}_{\rm lepton}$) \cite{Hooft}. Given $M^{}_\Delta
\sim {\cal O}(1)$ TeV, for example, this seesaw mechanism works to
generate $M^{}_\nu \sim {\cal O}(10^{-2})$ eV provided
$\lambda^{}_\Delta Y^{}_\Delta \sim {\cal O}(10^{-12})$ holds. The
neutrino mixing matrix $V$ is exactly unitary in the type-II
seesaw mechanism, simply because the heavy degrees of freedom do
not mix with the light ones.

There are totally seven physical Higgs bosons in the type-II
seesaw scheme: doubly-charged $H^{++}$ and $H^{--}$,
singly-charged $H^+$ and $H^-$, neutral $A^0$ (CP-odd), and
neutral $h^0$ and $H^0$ (CP-even), where $h^0$ is the SM-like
Higgs boson. Except for $M^2_{h^0}$, we get a quasi-degenerate
mass spectrum for other scalars \cite{T2SS}: $M^2_{H^{\pm \pm}} =
M^2_\Delta \approx M^2_{H^0} \approx M^2_{H^\pm} \approx
M^2_{A^0}$. As a consequence, the decay channels $H^{\pm \pm} \to
W^\pm H^\pm$ and $H^{\pm \pm} \to H^\pm H^\pm$ are kinematically
forbidden. The production of $H^{\pm\pm}$ at the LHC is mainly
through $q\bar{q} \to \gamma^*, Z^* \to H^{++}H^{--}$ and
$q\bar{q}^\prime \to W^* \to H^{\pm\pm}H^\mp$ processes, which do
not depend on the small Yukawa couplings.

The typical collider signatures in this seesaw scenario are the
lepton-number-violating $H^{\pm\pm} \to l^\pm_\alpha l^\pm_\beta$
decays as well as $H^+ \to l^+_\alpha \overline{\nu}$ and $H^- \to
l^-_\alpha \nu$ decays. Their branching ratios
\begin{eqnarray}
&& {\cal B}(H^{\pm\pm} \to l^\pm_\alpha l^\pm_\beta) \; = \;
\frac{ |(M^{}_\nu)^{}_{\alpha\beta}|^2}{\displaystyle
\sum_{\rho,\sigma} |(M^{}_\nu)^{}_{\rho\sigma}|^2} \left(2 -
\delta^{}_{\alpha\beta} \right) \; , \nonumber \\
&& {\cal B}(H^+ \to l^+_\alpha \overline{\nu}) \; = \;
\frac{\displaystyle \sum_\beta
|(M^{}_\nu)^{}_{\alpha\beta}|^2}{\displaystyle \sum_{\rho,\sigma}
|(M^{}_\nu)^{}_{\rho\sigma}|^2} \;
\end{eqnarray}
are closely related to the masses, flavor mixing angles and
CP-violating phases of three light neutrinos, because $M^{}_\nu =
V \widehat{M}^{}_\nu V^T$ with $\widehat{M}^{}_\nu = {\rm
Diag}\{m^{}_1, m^{}_2, m^{}_3\}$ holds. Some detailed analyses of
such decay modes together with the LHC signatures of $H^{\pm\pm}$
and $H^{\pm}$ bosons have been done in the literature \cite{Han2}.

It is worth pointing out that the following dimension-6 operator
can easily be derived from the type-II seesaw mechanism,
\begin{equation}
\frac{{\cal L}^{}_{\rm d=6}}{\Lambda^2} \; = \;
-\frac{\left(Y^{}_\Delta\right)^{}_{\alpha\beta}
\left(Y^{}_\Delta\right)^\dagger_{\rho\sigma}}{4 M^2_\Delta} ~
\left(\overline{l^{}_{\alpha \rm L}} \gamma^\mu l^{}_{\sigma \rm
L}\right) \left(\overline{l^{}_{\beta \rm L}} \gamma^{}_\mu
l^{}_{\rho \rm L}\right) \; ,
\end{equation}
which has two immediate low-energy effects: the non-standard
interactions of neutrinos and the lepton-flavor-violating
interactions of charged leptons. An analysis of such effects
provides us with some preliminary information \cite{Zhang08}:
\begin{itemize}
\item     The magnitudes of non-standard interactions of
neutrinos and the widths of lepton-flavor-violating tree-level
decays of charged leptons are both dependent on neutrino masses
$m^{}_i$ and flavor-mixing and CP-violating parameters of $V$.

\item     For a long-baseline neutrino oscillation experiment, the
neutrino beam encounters the earth matter and the electron-type
non-standard interaction contributes to the matter potential.

\item     At a neutrino factory, the lepton-flavor-violating
processes $\mu^-\rightarrow e^-\nu^{}_e\overline{\nu}^{}_\mu$ and
$\mu^+\rightarrow e^+\overline{\nu}_e\nu^{}_\mu$ could cause some
wrong-sign muons at a near detector.
\end{itemize}
Current experimental constraints tell us that such low-energy
effects are very small, but they might be experimentally
accessible in the future precision measurements.

\subsection{Type-(I+II) seesaw}

The type-(I+II) seesaw mechanism can be achieved by combining the
neutrino mass terms in Eqs. (1.4) and (1.5). After spontaneous
gauge symmetry breaking, we are left with the overall neutrino
mass term
\begin{equation}
-{\cal L}^{}_{\rm mass} \; = \; \frac{1}{2} ~ \overline{\left(
\nu^{}_{\rm L} ~N^{\rm c}_{\rm R}\right)} ~ \left( \begin{matrix}
M^{}_{\rm L} & M^{}_{\rm D} \cr M^T_{\rm D} & M^{}_{\rm R}
\end{matrix}\right) \left( \begin{matrix} \nu^{\rm c}_{\rm L} \cr N^{}_{\rm R}
\end{matrix} \right) + {\rm h.c.} \; ,
\end{equation}
where $M^{}_{\rm D} = Y^{}_\nu v/\sqrt{2}$ and $M^{}_{\rm L} =
Y^{}_\Delta v^{}_\Delta$ with $\langle H \rangle \equiv
v/\sqrt{2}$ and $\langle \Delta \rangle \equiv v^{}_\Delta$
corresponding to the vacuum expectation values of the neutral
components of the Higgs doublet $H$ and the Higgs triplet
$\Delta$. The $6\times 6$ neutrino mass matrix in Eq. (3.5) is
symmetric and can be diagonalized by the following unitary
transformation:
\begin{equation}
\left( \begin{matrix} V & R \cr S & U \end{matrix} \right)^\dagger
\left( \begin{matrix} M^{}_{\rm L} & M^{}_{\rm D} \cr M^T_{\rm D}
& M^{}_{\rm R} \end{matrix}\right) \left(\begin{matrix} V & R \cr
S & U \end{matrix} \right)^*  = \left( \begin{matrix}
\widehat{M}^{}_\nu & {\bf 0} \cr {\bf 0} & \widehat{M}^{}_N
\end{matrix}\right) \; ,
\end{equation}
where $\widehat{M}^{}_\nu = {\rm Diag}\{m^{}_1, m^{}_2, m^{}_3\}$
and $\widehat{M}^{}_N = {\rm Diag}\{M^{}_1, M^{}_2, M^{}_3\}$.
Needless to say, $V^\dagger V + S^\dagger S = VV^\dagger +
RR^\dagger = {\bf 1}$ holds as a consequence of the unitarity of
this transformation. Hence $V$, the flavor mixing matrix of three
light Majorana neutrinos, must be non-unitary if $R$ and $S$ are
non-zero.

In the leading-order approximation, the type-(I+II) seesaw formula
reads as
\begin{equation}
M^{}_\nu \; = \; M^{}_{\rm L} - M^{}_{\rm D} M^{-1}_{\rm R}
M^T_{\rm D} \; .
\end{equation}
Hence type-I and type-II seesaws can be regarded as two extreme
cases of the type-(I+II) seesaw. Note that two mass terms in Eq.
(3.7) are possibly comparable in magnitude. If both of them are
small, their contributions to $M^{}_\nu$ may have significant
interference effects which make it practically impossible to
distinguish between type-II and type-(I+II) seesaws \cite{Ren};
but if both of them are large, their contributions to $M^{}_\nu$
must be destructive. The latter case unnaturally requires a
significant cancellation between two big quantities in order to
obtain a small quantity, but it is interesting in the sense that
it may give rise to possibly observable collider signatures of
heavy Majorana neutrinos \cite{Chao}.

Let me briefly describe a particular type-(I+II) seesaw model and
comment on its possible LHC signatures. First, we assume that both
$M^{}_i$ and $M^{}_\Delta$ are of ${\cal O}(1)$ TeV. Then the
production of $H^{\pm\pm}$ and $H^\pm$ bosons at the LHC is
guaranteed, and their lepton-number-violating signatures will
probe the Higgs triplet sector of the type-(I+II) seesaw
mechanism. On the other hand, ${\cal O}(M^{}_{\rm D}/M^{}_{\rm R})
\lesssim {\cal O}(0.1)$ is possible as a result of ${\cal O}(
M^{}_{\rm R}) \sim {\cal O}(1)$ TeV and ${\cal O}(M^{}_{\rm D})
\lesssim {\cal O}(v)$, such that appreciable signatures of
$N^{}_i$ can be achieved at the LHC. Second, the small mass scale
of $M^{}_\nu$ implies that the relation ${\cal O}(M^{}_{\rm L})
\sim {\cal O}(M^{}_{\rm D} M^{-1}_{\rm R} M^T_{\rm D})$ must hold.
In other words, it is the significant but incomplete cancellation
between $M^{}_{\rm L}$ and $M^{}_{\rm D} M^{-1}_{\rm R} M^T_{\rm
D}$ terms that results in the non-vanishing but tiny masses for
three light neutrinos. We admit that dangerous radiative
corrections to two mass terms of $M^{}_\nu$ require a delicate
fine-tuning of the cancellation at the loop level \cite{Si}. But
this scenario allows us to reconstruct $M^{}_{\rm L}$ via the
excellent approximation $M^{}_{\rm L} = V \widehat{M}^{}_\nu V^T +
R \widehat{M}^{}_N R^T \approx R \widehat{M}^{}_N R^T$, such that
the elements of the Yukawa coupling matrix $Y^{}_\Delta$ read
\begin{equation}
\left(Y^{}_\Delta\right)^{}_{\alpha \beta} \; = \;
\frac{\left(M^{}_{\rm L}\right)^{}_{\alpha \beta}}{v^{}_\Delta}
\approx \sum^3_{i=1} \frac{R^{}_{\alpha i} R^{}_{\beta i}
M^{}_i}{v^{}_\Delta} \; ,
\end{equation}
where the subscripts $\alpha$ and $\beta$ run over $e$, $\mu$ and
$\tau$. This result implies that the leptonic decays of $H^{\pm
\pm}$ and $H^\pm$ bosons depend on both $R$ and $M^{}_i$, which
actually determine the production and decays of $N^{}_i$. Thus we
have established an interesting correlation between the singly- or
doubly-charged Higgs bosons and the heavy Majorana neutrinos. To
observe the correlative signatures of $H^\pm$, $H^{\pm\pm}$ and
$N^{}_i$ at the LHC will serve for a direct test of this
type-(I+II) seesaw model.

To illustrate, here I focus on the {\it minimal} type-(I+II)
seesaw model with a single heavy Majorana neutrino, where $R$ can
be parametrized in terms of three rotation angles $\theta^{}_{i4}$
and three phase angles $\delta^{}_{i4}$ (for $i=1,2,3$)
\cite{Xing08}. In this case, we have
\begin{eqnarray}
\omega^{}_1 & \equiv & \frac{\sigma (pp \to \mu^+ \mu^+ W^-
X)|^{}_{N^{}_1}}{\sigma (pp \to \mu^+ \mu^+ H^- X)|^{}_{H^{++}}}
\; \approx \; \frac{\sigma^{}_N}{\sigma^{}_H} \cdot \frac{s^2_{14}
+ s^2_{24} + s^2_{34}}{4} \; , \nonumber \\
\omega^{}_2 & \equiv & \frac{\sigma (pp \to \mu^+ \mu^+ W^-
X)|^{}_{N^{}_1}}{\sigma (pp \to \mu^+ \mu^+ H^{--}
X)|^{}_{H^{++}}} \; \approx \; \frac{\sigma^{}_N}{\sigma^{}_{\rm
pair}} \cdot \frac{s^2_{14} + s^2_{24} + s^2_{34}}{4} \;
\end{eqnarray}
for $s^{}_{i4} \equiv \sin\theta^{}_{i4} \lesssim {\cal O}(0.1)$,
where $\sigma^{}_{N} \equiv \sigma(pp \to l^+_\alpha N^{}_1
X)/|R^{}_{\alpha 1}|^2$, $\sigma^{}_{H} \equiv \sigma(pp \to
H^{++} H^- X)$ and $\sigma^{}_{\rm pair} \equiv \sigma(pp \to
H^{++} H^{--}X)$ are three reduced cross sections \cite{Si}. Here
let me omit a numerical illustration of $\omega^{}_1$ and
$\omega^{}_2$ changing with $M^{}_1$ at the LHC with an integrated
luminosity of $300 ~ {\rm fb}^{-1}$, although it may give one a
ball-park feeling of possible collider signatures of $N^{}_1$ and
$H^{\pm\pm}$ and their correlation \cite{Si}.

\subsection{Type-III seesaw}

The lepton mass terms in the type-III seesaw scheme have already
been given in Eq. (1.7). After spontaneous gauge symmetry breaking,
we are left with
\begin{eqnarray}
-{\cal L}^{}_{\rm mass} & = & \frac{1}{2} ~\overline{\left(
\nu^{}_{\rm L} ~ \Sigma^0 \right)} \left( \begin{matrix} {\bf 0} &
M^{}_{\rm D} \cr M^T_{\rm D} & M^{}_\Sigma \cr \end{matrix} \right)
\left(
\begin{matrix} \nu^{\rm c}_{\rm L} \cr {\Sigma^{0}}^{\rm c} \cr \end{matrix} \right)
+ {\rm h.c.} \; , \nonumber \\
-{\cal L}^\prime_{\rm mass} & = & \overline{\left( e^{}_{\rm L} ~
\Psi^{}_{\rm L} \right)} \left( \begin{matrix} M^{}_l & \sqrt{2}
M^{}_{\rm D} \cr {\bf 0} & M^{}_\Sigma \cr \end{matrix} \right)
\left( \begin{matrix} E^{}_{\rm R} \cr \Psi^{}_{\rm R} \cr
\end{matrix} \right) + {\rm h.c.} \; ,
\end{eqnarray}
respectively, for neutral and charged fermions, where $M^{}_l =
Y^{}_l v/\sqrt{2}~$, $M^{}_{\rm D} = Y^{}_\Sigma v/\sqrt{2}~$, and
$\Psi = \Sigma^- + {\Sigma^+}^{\rm c}$. The symmetric $6\times 6$
neutrino mass matrix can be diagonalized by the following unitary
transformation:
\begin{equation}
\left( \begin{matrix} V & R \cr S & U \cr \end{matrix}
\right)^\dagger \left( \begin{matrix} {\bf 0} & M^{}_{\rm D} \cr
M^T_{\rm D} & M^{}_\Sigma \cr \end{matrix} \right) \left(
\begin{matrix} V & R \cr S & U \cr \end{matrix} \right)^* = \left(
\begin{matrix} \widehat{M}^{}_\nu & {\bf 0} \cr {\bf 0} &
\widehat{M}^{}_\Sigma \cr \end{matrix} \right) \; ,
\end{equation}
where $\widehat{M}^{}_\nu = {\rm Diag}\{m^{}_1, m^{}_2, m^{}_3 \}$
and $\widehat{M}^{}_\Sigma = {\rm Diag}\{M^{}_1, M^{}_2, M^{}_3 \}$.
In the leading-order approximation, this diagonalization yields the
type-III seesaw formula $M^{}_\nu = -M^{}_{\rm D} M^{-1}_\Sigma
M^T_{\rm D}$, which is equivalent to the one derived from the
effective dimension-5 operator in Eq. (1.11). Let us use one
sentence to comment on the similarities and differences between
type-I and type-III seesaw mechanisms \cite{Abada}: the
non-unitarity of the $3\times 3$ neutrino mixing matrix $V$ has
appeared in both cases, although the modified couplings between the
$Z^0$ boson and three light neutrinos differ and the non-unitary
flavor mixing is also present in the couplings between the $Z^0$
boson and three charged leptons in the type-III seesaw scenario.

At the LHC, the typical lepton-number-violating signatures of the
type-III seesaw mechanism can be $pp \to \Sigma^+ \Sigma^0 \to
l^+_\alpha l^+_\beta + Z^0W^-(\to 4j)$ and $pp \to \Sigma^-
\Sigma^0 \to l^-_\alpha l^-_\beta + Z^0W^+(\to 4j)$ processes. A
detailed analysis of such collider signatures have been done in
the literature \cite{Hambye}. As for the low-energy phenomenology,
a consequence of this seesaw scenario is the non-unitarity of the
$3\times 3$ flavor mixing matrix $N$ ($\approx V$) in both
charged- and neutral-current interactions \cite{Abada}. Current
experimental bounds on the deviation of $NN^\dagger$ from the
identity matrix are at the $0.1\%$ level, much stronger than those
obtained in the type-I seesaw scheme, just because the
flavor-changing processes with charged leptons are allowed at the
tree level in the type-III seesaw mechanism.

I like to mention that an interesting type-(I+III) seesaw model has
recently been proposed \cite{BS}, and its phenomenological and
cosmological consequences together with its possible collider
signatures have also been explored \cite{BS2}.

\subsection{Double (inverse) seesaw}

Given the naturalness and testability as two prerequisites, the
double or inverse seesaw mechanism \cite{MV} is another
interesting possibility of generating tiny neutrino masses at the
TeV scale. The idea of this seesaw picture is to add three heavy
right-handed neutrinos $N^{}_{\rm R}$, three SM gauge-singlet
neutrinos $S^{}_{\rm R}$ and one Higgs singlet $\Phi$ into the SM,
such that the gauge-invariant lepton mass terms can be written as
\begin{equation}
-{\cal L}^{}_{\rm lepton} \; =\; \overline{l^{}_{\rm L}} Y^{}_l H
E^{}_{\rm R} + \overline{l^{}_{\rm L}} Y^{}_\nu \tilde{H} N^{}_{\rm
R} + \overline{N^{\rm c}_{\rm R}} Y^{}_S \Phi S^{}_{\rm R} +
\frac{1}{2} \overline{S^{\rm c}_{\rm R}} \mu S^{}_{\rm R} + {\rm
h.c.} \; ,
\end{equation}
where the $\mu$-term is naturally small according to t' Hooft's
naturalness criterion \cite{Hooft}, because it violates the lepton
number. After spontaneous gauge symmetry breaking, the overall
neutrino mass term turns out to be
\begin{equation}
-{\cal L}^{}_{\rm mass} \; =\; \frac{1}{2}
~\overline{\left(\nu^{}_{\rm L} ~N^{\rm c}_{\rm R} ~S^{\rm c}_{\rm
R}\right)} \left( \begin{matrix} {\bf 0} & M^{}_{\rm D} & {\bf 0}
\cr M^T_{\rm D} & {\bf 0} & M^{}_S \cr {\bf 0} & M^T_S & \mu \cr
\end{matrix} \right) \left( \begin{matrix} \nu^{\rm c}_{\rm L} \cr
N^{}_{\rm R} \cr S^{}_{\rm R} \cr \end{matrix} \right) \; ,
\end{equation}
where $M^{}_{\rm D} = Y^{}_\nu \langle H\rangle$ and $M^{}_S =
Y^{}_S \langle \Phi\rangle$. A diagonalization of the symmetric
$9\times 9$ matrix in Eq. (3.13) leads us to the effective light
neutrino mass matrix
\begin{equation}
M^{}_\nu \; = \; M^{}_{\rm D} \frac{1}{M^T_S} \mu \frac{1}{M^{}_S}
M^T_{\rm D} \;
\end{equation}
in the leading-order approximation. Hence the smallness of
$M^{}_\nu$ can be attributed to both the smallness of $\mu$ itself
and the doubly-suppressed $M^{}_{\rm D}/M^{}_S$ term for
$M^{}_{\rm D} \ll M^{}_S$. For example, $\mu \sim {\cal O}(1)$ keV
and $M^{}_{\rm D}/M^{}_S \sim {\cal O}(10^{-2})$ naturally give
rise to a sub-eV $M^{}_\nu$. One has $M^{}_\nu = {\bf 0}$ in the
limit $\mu \rightarrow {\bf 0}$, which reflects the restoration of
the slightly-broken lepton number. The heavy sector consists of
three pairs of pseudo-Dirac neutrinos whose CP-conjugated Majorana
components have a tiny mass splitting characterized by the order
of $\mu$.

A minimal inverse seesaw scenario, in which only two pairs of the
gauge-singlet neutrinos $N^{}_{\rm R}$ and $S^{}_{\rm R}$ are
introduced, has recently been proposed \cite{MISS}. Its LHC
signatures and low-energy consequences deserve some further
studies.

\section{Non-unitary neutrino mixing and CP violation}

It is worth remarking that the charged-current interactions of light
and heavy Majorana neutrinos are not completely independent in
either the type-I seesaw or the type-(I+II) seesaw. The standard
charged-current interactions of $\nu^{}_i$ and $N^{}_i$ are
\begin{equation}
-{\cal L}^{}_{\rm cc} = \frac{g}{\sqrt{2}} \overline{\left(e~ \mu~
\tau\right)^{}_{\rm L}} ~\gamma^\mu \left[ V \left(
\begin{matrix} \nu^{}_1 \cr \nu^{}_2 \cr \nu^{}_3
\end{matrix} \right)^{}_{\rm L} + R \left(
\begin{matrix} N^{}_1 \cr N^{}_2 \cr N^{}_3 \end{matrix}
\right)^{}_{\rm L} \right] W^-_\mu + {\rm h.c.} \; ,
\end{equation}
where $V$ is just the light neutrino mixing matrix responsible for
neutrino oscillations, and $R$ describes the strength of
charged-current interactions between $(e, \mu, \tau)$ and
$(N^{}_1, N^{}_2, N^{}_3)$. Since $V$ and $R$ belong to the same
unitary transformation done in Eq. (3.6), they must be correlated
with each other and their correlation signifies an important
relationship between neutrino physics and collider physics.

It has been shown that $V$ and $R$ share nine rotation angles
($\theta^{}_{i4}$, $\theta^{}_{i5}$ and $\theta^{}_{i6}$ for $i=1$,
$2$ and $3$) and nine phase angles ($\delta^{}_{i4}$,
$\delta^{}_{i5}$ and $\delta^{}_{i6}$ for $i=1$, $2$ and $3$)
\cite{Xing08}. To see this point clearly, let me decompose $V$ into
$V = A V^{}_0$, where
\begin{equation}
V^{}_0 = \left( \begin{matrix} c^{}_{12} c^{}_{13} &
\hat{s}^*_{12} c^{}_{13} & \hat{s}^*_{13} \cr -\hat{s}^{}_{12}
c^{}_{23} - c^{}_{12} \hat{s}^{}_{13} \hat{s}^*_{23} & c^{}_{12}
c^{}_{23} - \hat{s}^*_{12} \hat{s}^{}_{13} \hat{s}^*_{23} &
c^{}_{13} \hat{s}^*_{23} \cr \hat{s}^{}_{12} \hat{s}^{}_{23} -
c^{}_{12} \hat{s}^{}_{13} c^{}_{23} & -c^{}_{12} \hat{s}^{}_{23} -
\hat{s}^*_{12} \hat{s}^{}_{13} c^{}_{23} & c^{}_{13} c^{}_{23} \cr
\end{matrix} \right)
\end{equation}
with $c^{}_{ij} \equiv \cos\theta^{}_{ij}$ and $\hat{s}^{}_{ij}
\equiv e^{i\delta^{}_{ij}} \sin\theta^{}_{ij}$ is just the standard
parametrization of the $3\times 3$ unitary neutrino mixing matrix
(up to some proper phase rearrangements) \cite{PDG,Xing04}. Because
of $VV^\dagger = AA^\dagger = {\bf 1} - RR^\dagger$, it is obvious
that $V \rightarrow V^{}_0$ in the limit of $A \rightarrow {\bf 1}$
(or equivalently, $R \rightarrow {\bf 0}$). Considering the fact
that the non-unitarity of $V$ must be a small effect (at most at the
percent level as constrained by current neutrino oscillation data
and precision electroweak data \cite{Antusch}), we expect $s^{}_{ij}
\lesssim {\cal O}(0.1)$ (for $i=1,2,3$ and $j=4,5,6$) to hold. Then
we obtain \cite{Xing08}
\begin{eqnarray}
A & = & {\bf 1} - \left( \begin{matrix} \frac{1}{2} \left(
s^2_{14} + s^2_{15} + s^2_{16} \right) & 0 & 0 \cr \hat{s}^{}_{14}
\hat{s}^*_{24} + \hat{s}^{}_{15} \hat{s}^*_{25} + \hat{s}^{}_{16}
\hat{s}^*_{26} & \frac{1}{2} \left( s^2_{24} + s^2_{25} + s^2_{26}
\right) & 0 \cr \hat{s}^{}_{14} \hat{s}^*_{34} + \hat{s}^{}_{15}
\hat{s}^*_{35} + \hat{s}^{}_{16} \hat{s}^*_{36} & \hat{s}^{}_{24}
\hat{s}^*_{34} + \hat{s}^{}_{25} \hat{s}^*_{35} + \hat{s}^{}_{26}
\hat{s}^*_{36} & \frac{1}{2} \left( s^2_{34} +
s^2_{35} + s^2_{36} \right) \cr \end{matrix} \right) , \nonumber \\
R & = & {\bf 0} + \left( \begin{matrix} \hat{s}^*_{14} &
\hat{s}^*_{15} & \hat{s}^*_{16} \cr \hat{s}^*_{24} &
\hat{s}^*_{25} & \hat{s}^*_{26} \cr \hat{s}^*_{34} &
\hat{s}^*_{35} & \hat{s}^*_{36} \cr \end{matrix} \right)
\end{eqnarray}
as two excellent approximations. A striking consequence of the
non-unitarity of $V$ is the loss of universality for the Jarlskog
invariants of CP violation \cite{J}, $J^{ij}_{\alpha\beta} \equiv
{\rm Im}(V^{}_{\alpha i} V^{}_{\beta j} V^*_{\alpha j} V^*_{\beta
i})$, where the Greek indices run over $(e, \mu, \tau)$ and the
Latin indices run over $(1,2,3$). For example, the extra
CP-violating phases of $V$ are possible to give rise to a
significant asymmetry between $\nu^{}_\mu \rightarrow \nu^{}_\tau$
and $\overline{\nu}^{}_\mu \rightarrow \overline{\nu}^{}_\tau$
oscillations.

The probability of $\nu^{}_\alpha \rightarrow \nu^{}_\beta$
oscillations in vacuum, defined as $P^{}_{\alpha\beta}$, is given by
\cite{Xing08}
\begin{equation}
P^{}_{\alpha\beta} \; = \frac{\displaystyle \sum^{}_i |V^{}_{\alpha
i}|^2 |V^{}_{\beta i}|^2 + 2 \sum^{}_{i<j} {\rm Re} \left(
V^{}_{\alpha i} V^{}_{\beta j} V^*_{\alpha j} V^*_{\beta i} \right)
\cos \Delta^{}_{ij} - 2 \sum^{}_{i<j} J^{ij}_{\alpha\beta}
\sin\Delta^{}_{ij}}{\displaystyle \left(
VV^\dagger\right)^{}_{\alpha\alpha} \left(
VV^\dagger\right)^{}_{\beta\beta}} \; ,
\end{equation}
where $\Delta^{}_{ij} \equiv \Delta m^2_{ij} L/(2E)$ with $\Delta
m^2_{ij} \equiv m^2_i - m^2_j$, $E$ being the neutrino beam energy
and $L$ being the baseline length. If $V$ is exactly unitary (i.e.,
$A = {\bf 1}$ and $V = V^{}_0$), the denominator of Eq. (4.4) will
become unity and the conventional formula of $P^{}_{\alpha\beta}$
will be reproduced. Note that $\nu^{}_\mu \rightarrow \nu^{}_\tau$
and $\overline{\nu}^{}_\mu \rightarrow \overline{\nu}^{}_\tau$
oscillations may serve as a good tool to probe possible signatures
of non-unitary CP violation \cite{Xing08,Yasuda}. To illustrate this
point, we consider a short- or medium-baseline neutrino oscillation
experiment with $|\sin\Delta^{}_{13}| \sim |\sin\Delta^{}_{23}| \gg
|\sin\Delta^{}_{12}|$, in which the terrestrial matter effects are
expected to be insignificant or negligibly small. Then the dominant
CP-conserving and CP-violating terms of $P(\nu^{}_\mu \rightarrow
\nu^{}_\tau)$ and $P(\overline{\nu}^{}_\mu \rightarrow
\overline{\nu}^{}_\tau)$ are
\begin{eqnarray}
P(\nu^{}_\mu \rightarrow \nu^{}_\tau) & \approx & \sin^2
2\theta^{}_{23} \sin^2 \frac{\Delta^{}_{23}}{2} ~ - ~ 2 \left(
J^{23}_{\mu\tau} + J^{13}_{\mu\tau} \right) \sin\Delta^{}_{23} \;
, \nonumber \\
P(\overline{\nu}^{}_\mu \rightarrow \overline{\nu}^{}_\tau) &
\approx & \sin^2 2\theta^{}_{23} \sin^2 \frac{\Delta^{}_{23}}{2} ~
+ ~ 2 \left( J^{23}_{\mu\tau} + J^{13}_{\mu\tau} \right)
\sin\Delta^{}_{23} \; ,
\end{eqnarray}
where the good approximation $\Delta^{}_{13} \approx
\Delta^{}_{23}$ has been used in view of the experimental fact
$|\Delta m^2_{13}| \approx |\Delta m^2_{23}| \gg |\Delta
m^2_{12}|$, and the sub-leading and CP-conserving ``zero-distance"
effect \cite{Antusch} has been omitted. For simplicity, I take
$V^{}_0$ to be the exactly tri-bimaximal mixing pattern \cite{TB}
(i.e., $\theta^{}_{12} = \arctan(1/\sqrt{2})$, $\theta^{}_{13} =0$
and $\theta^{}_{23} =\pi/4$ as well as $\delta^{}_{12} =
\delta^{}_{13} = \delta^{}_{23} =0$) and then arrive at
\cite{Xing08}
\begin{equation}
2\left( J^{23}_{\mu\tau} + J^{13}_{\mu\tau} \right) \; \approx \;
\sum^6_{l=4} s^{}_{2l} s^{}_{3l} \sin \left( \delta^{}_{2l} -
\delta^{}_{3l} \right) \; .
\end{equation}
Given $s^{}_{2l} \sim s^{}_{3l} \sim {\cal O}(0.1)$ and
$(\delta^{}_{2l} - \delta^{}_{3l}) \sim {\cal O}(1)$ (for
$l=4,5,6$), this non-trivial CP-violating quantity can reach the
percent level. When a long-baseline neutrino oscillation experiment
is concerned, however, the terrestrial matter effects must be taken
into account because they might fake the genuine CP-violating
signals \cite{Xing00}. As for $\nu^{}_\mu \rightarrow \nu^{}_\tau$
and $\overline{\nu}^{}_\mu \rightarrow \overline{\nu}^{}_\tau$
oscillations under discussion, the dominant matter effect results
from the neutral-current interactions and modifies the CP-violating
quantity of Eq. (4.6) in the following way \cite{Xing09}:
\begin{eqnarray}
2\left( J^{23}_{\mu\tau} + J^{13}_{\mu\tau} \right) \;
\Longrightarrow \; \sum^6_{l=4} s^{}_{2l} s^{}_{3l} \left[ \sin
\left( \delta^{}_{2l} - \delta^{}_{3l} \right) + A^{}_{\rm NC} L
\cos \left( \delta^{}_{2l} - \delta^{}_{3l} \right) \right] \; ,
\end{eqnarray}
where $A^{}_{\rm NC} = G^{}_{\rm F} N^{}_n /\sqrt{2}~$ with
$N^{}_n$ being the background density of neutrons, and $L$ is the
baseline length. It is easy to find $A^{}_{\rm NC} L \sim {\cal
O}(1)$ for $L \sim 4 \times 10^3$ km.

\section{Inconclusive concluding remarks}

Although the seesaw ideas are elegant, they have to appeal for
some or many new degrees of freedom in order to interpret the
observed neutrino mass hierarchy and lepton flavor mixing.
According to Weinberg's {\it third law of progress in theoretical
physics} \cite{Weinberg80}, ``you may use any degrees of freedom
you like to describe a physical system, but if you use the wrong
ones, you will be sorry." What could be better?

Anyway, we hope that the LHC might open a new window for us to
understand the origin of neutrino masses and the dynamics of lepton
number violation. A TeV seesaw might work ({\it naturalness}?) and
its heavy degrees of freedom might show up at the LHC ({\it
testability}?). A bridge between collider physics and neutrino
physics is highly anticipated and, if it exists, will lead to rich
phenomenology.

\vspace{0.5cm}

I am indebted to T. Kobayashi and other organizers of this symposium
for kind invitation and warm hospitality. I am also grateful to S.
Zhou and W. Chao for many useful discussions. This work is supported
in part by the National Natural Science Foundation of China under
grant No. 10425522 and No. 10875131.

\end{document}